\documentclass[aps,prl,twocolumn]{revtex4}
\usepackage{amssymb,graphicx}
\usepackage{times}
\newcommand{\fig}[2]{\includegraphics[width=#1]{./figures/#2}}

\newlength{\bilderlength}

\begin{document}
\bibliographystyle{KAY}
%%%%%%%%%%%%%%%%%%%%%%%%%%%%%%%%%%%%%%%%%%%%%%%%%%%%%%%%%%%%%%%%%%%%%%%%%%%%
\title{
%\sffamily \bfseries \Large
Measuring functional
renormalization group fixed-point functions for pinned manifolds}

\author{
%\sffamily\bfseries\normalsize
A. Alan Middleton$^1$,
Pierre Le Doussal$^2$, and Kay J\"org Wiese$^2$ \vspace*{3mm} }

\affiliation{$^1$Department of Physics, Syracuse University, Syracuse, NY
13244, USA.\\
$^2$CNRS-Laboratoire de Physique Th{\'e}orique de l'Ecole
Normale Sup{\'e}rieure, 24 rue Lhomond, 75005 Paris, France. }

\date{\small\today}

\begin{abstract}
Exact numerical minimization of interface energies is used to test the
functional renormalization group (FRG) analysis for interfaces pinned
by quenched disorder.  The fixed-point function $R(u)$ (the correlator
of the coarse-grained disorder) is computed.  In dimensions $D=d+1$, a
linear cusp in $R''(u)$ is confirmed for random bond ($d=1,2,3$),
random field ($d=0,2,3$), and periodic ($d=2,3$) disorders. The
functional shocks that lead to this cusp are seen. Small, but
significant, deviations from 1-loop FRG results are compared to 2-loop
corrections. The cross-correlation for two copies of disorder is compared
with a recent FRG study of chaos.
\end{abstract}
\maketitle

Systems with quenched (frozen-in) disorder often exhibit glassy phases
at low temperature.  Standard perturbative methods fail to describe
these phases and exact results are limited to 1D and mean field models
\cite{HuseHenleyFisher,Johansson2000,MP}.  It has been quite a
challenge to develop field theoretic and renormalization group (RG)
methods, which must include both multiple metastable states and
spatial fluctuations in finite dimensions, to describe universal
properties of these phases.  Proposed field theories are
unconventional and harder to control than those developed for pure
critical systems. An expansion around the mean-field replica symmetry
(and ergodicity) broken (RSB) solution, much studied in spin glasses,
is very difficult even at the 1-loop level \cite{Dominicis}. The
functional RG (FRG) was developed for elastic objects pinned by
substrate disorder and random fields. This class has numerous physical
realizations, including vortex lattices, magnetic systems, and charge
density waves
\cite{DSFisher1986,reviews_pinning,GiamarchiLeDoussal1995,NattermannBookYoung}.
The 1-loop FRG has been extended to describe, e.g., depinning of a
driven interface \cite{depinning}, activated dynamics \cite{creep},
quantum models \cite{quantum}, and sensitivity of configurations to
disorder changes (``chaos'') \cite{chaospld}. Since the FRG
parameterizes the effective action by functions, rather than the few
couplings of standard RG, it is better suited to handle an infinite
number of marginal parameters at the upper critical dimension (or
runaway flows as in correlated fermions \cite{FRGfermions}).

When applying the FRG to pinned elastic manifolds parameterized by a
scalar displacement field $u(x)$, the function in the effective action
whose flow is relevant below $d=4$ is denoted by $R(u)$.  Physically,
this function represents a coarse graining of the correlator of the
pinning potential; it encodes an infinite number of couplings,
$R^{(2n)}(0)$, $n=0\ldots\infty$.
An unusual feature of the theory is
that $R''(u)$ can develop a linear cusp around $u=0$ at finite scale
\cite{DSFisher1986}.
In the space of non-analytic functions,
perturbative control was recovered to one-loop order (i.e., to
$O(\epsilon=4-d)$) and fixed-point functions $R(u)$ obtained for
various universality classes \cite{DSFisher1986,BalentsFisher1993,
GiamarchiLeDoussal1995}.
The relations between this cusp singularity,
multiple metastable states and shocks in energy landscapes have been
vividly described
\cite{BalentsBouchaudMezard1996}. The FRG agrees with phenomenological
models and successfully predicts the roughness exponent $\zeta$ of the
pinned interface, with the disorder-averaged correlation function
$\overline{(u(x)-u(0))^2} \sim x^{2\zeta}$
\cite{Middleton1995,books,NohRieger2001}.

Though much has been achieved, it has been questioned
\cite{BalentsFisher1993} whether the FRG can be extended in a systematic
loop expansion, i.e., to higher order in $\epsilon$. Dealing with a
non-analytic action is very subtle \cite{ChauveLeDoussalWiese2000a},
and even 1-loop consistency is not obvious \cite{exactRG}.
Recently, a candidate renormalizable field theory for statics
\cite{ChauveLeDoussalWiese2000a,LeDoussalWieseChauve2004} (and a
distinct one for depinning \cite{LeDoussalWieseChauve2002}) was
obtained beyond one loop. Crucial to its construction is the
property {\it that the cusp remains linear to higher orders}. If
confirmed, the FRG provides a simpler method to attack glass
problems where the RSB phenomenology can be avoided.

\newlength{\figsize}
\setlength{\figsize}{1\columnwidth}
\begin{figure}[b]
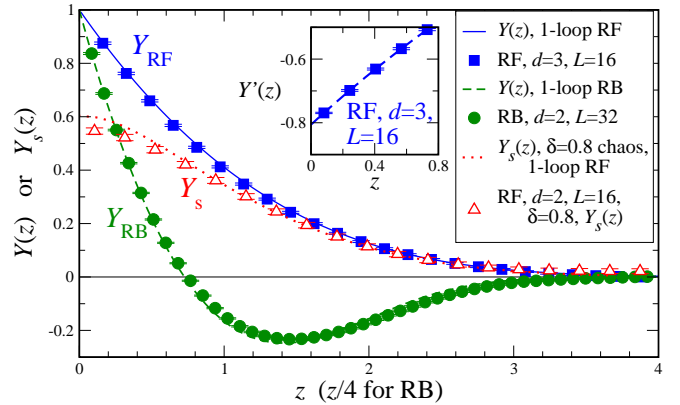
\setlength{\unitlength}{1.4mm}
\fboxsep0mm \mbox{\fig{\figsize}{compareRFRBchaos}} \caption{Filled
symbols show numerical
results for $Y(z)$, a normalized form of the interface displacement
correlator $-R''(u)$ [Eq.\
(\ref{split})], for $D=2+1$ random field (RF) and $D=3+1$ random
bond (RB) disorders. These suggest a linear
cusp. The inset plots
the numerical derivative $Y'(z)$, with intercept
$Y'(0)\approx -0.807$ from a quadratic fit
(dashed line).
Open symbols plot the cross-correlator ratio
$Y_s(z)=\Delta_{12}(z)/\Delta_{11}(0)$ between two related copies of
RF disorder. It does not exhibit a cusp.
The points are for confining wells with width given
by $M^2=0.02$.
Comparisons to 1-loop FRG predictions (curves) are made with no
adjustable parameters.
\label{figure1}}
\end{figure}

\begin{figure}[t]
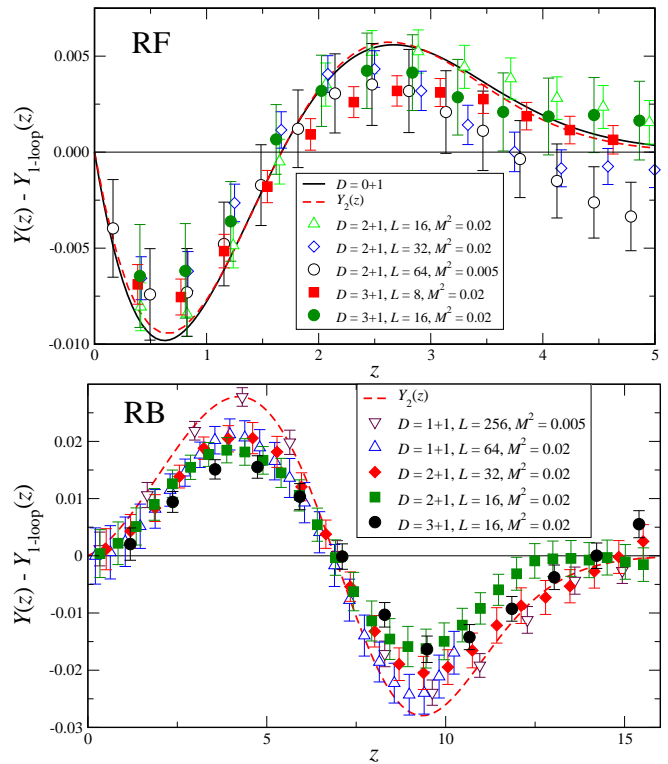
\setlength{\unitlength}{1.4mm}
\fboxsep0mm \mbox{\fig{\figsize}{RF_residualswith64}}
\mbox{\fig{\figsize}{RB_residuals}} \caption{The
difference between the normalized correlator $Y(z)$ and the 1-loop
prediction $Y_{1}(z)$ for RF disorder in $D=(1,2)+1$ and RB
disorder in $D=(1,2,3)+1$. The dashed lines
are the linear 2-loop correction $Y_2(z)=\frac{d Y(z)}{d
\epsilon}|_{\epsilon=0}$ of
 Eq.~(\ref{expansion}). For each disorder class, the
data are close to each other and to the $d=0$ and $\epsilon=1$ linear two-loop
estimates, but are distinct from the 1-loop result.
} \label{RFRBresiduals}
\end{figure}

\begin{figure}[t]
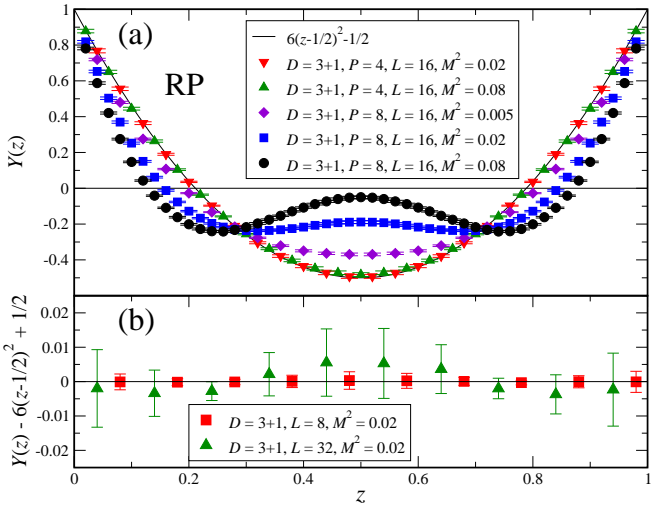
\setlength{\unitlength}{1.4mm}
\fboxsep0mm \mbox{\fig{\figsize}{RP_residuals}} \caption{(a)
Plots of the normalized pinning force correlator $Y(z)$ for RP
disorder in $d=3$. For these values of $m$, the period $P=4$
points have mostly converged to the parabolic RP fixed-point
function, while the $P=8$ curves are still crossing over from the RB to
the RP universality class. (b) Residuals relative to
the parabolic shape vanish, within error bars, for larger sizes
and $P=4$.} \label{figureRP}
\end{figure}

This paper presents a new level of ``smoking gun'' tests of the FRG
for manifolds, by {\it directly measuring} the fixed-point function
$R(u)$ for three universality classes
(Figs.~\ref{figure1}-\ref{figureRP}). This is achieved, as proposed in
Ref.\ \cite{pld}, by adding to the disorder a parabolic potential
(i.e., a mass $m$) with a variable minimum location $v$.  The
resulting sample-dependent free energy $\hat V(v)$ defines a
renormalized potential whose second cumulant correlator in $v$ space
is {\it the same} $R(v)$ function as defined in the field theory (from
the replicated effective action - deviations arise only in higher
cumulants \cite{pld}). This is analogous to measuring the coupling
constant and the distribution of total magnetization in pure systems,
which underlie phenomenological RG and finite size scaling
\cite{Binder}. As in pure systems, the FRG predictions are universal
at coarse grained scales, but require specifying the large scale
BCs. The mass provides these conditions and also allows one to control
and quantify the zero mode (center of mass) fluctuations, yielding the
coupling function $R(u)$. The same procedure allowed an exact
calculation \cite{pld} of $R(v)$ for the $D=0+1$ theory with RF
disorder (Sinai's model).

We numerically compute the FRG zero temperature fixed-point functions
using exact ground state configurations.  We study interfaces embedded
in dimensions $D=d+1$, $d=0,1,2,3$, including random bond (RB), random
field (RF), and periodic (RP) disorder universality classes. We focus
on universal, parameter free functions; treatment of universal
amplitudes requires more details and is presented separately
\cite{us_future}. The linear cusp in $\Delta(u)=-R''(u)$ is confirmed
in all cases. For periodic disorder, $\Delta(u)$ is consistent with
the conjectured parabolic form.  For RB and RF disorder, the scaled
$\Delta(u)$ are distinct from the 1-loop calculations and are closer
to the two-loop results, though the curves exhibit at most a weak
dependence on $d$.  The functional shocks responsible for the cusp in
$\Delta(u)$ are directly seen.  The higher statistics of these shocks
are consistent with $d=0$ Burgers intermittency. Cross-correlation
(chaos) fixed points for two related copies of the disorder show a
rounding of $\Delta(u)$ that is consistent with recent FRG predictions
\cite{chaospld}.

The continuum Hamiltonian for an interface $u(x)$ of internal area
$\Omega$ with elastic constant $K$, confined in a parabolic well
centered at $v$ is
\begin{equation}
{\cal H}(v) =\int_{\Omega}d^dx\,\left\{\frac{K}{2}\left(\nabla
u \right)^{2} +\frac{m^{2}}{2} (u - v)^{2}+V[x,u(x)]\right\}
\label{eq:H_cont}
\end{equation}
where the random potential $V$ has correlations $\overline{V(0,x)
V(u,x')}= R_0(u) \delta^{(d)}(x-x')$.  The RB universality class
describes short ranged $R_0(u)$, the RF class has $R_0(u) \sim -
\sigma |u|$ at large $u$, while the RP class describes periodic
correlations, e.g., $R_0(u+1)=R_0(u)$. The bare correlator $R_0(u)$
becomes $R(u)$ upon coarse graining. Given a UV cutoff scale $b$,
for fixed $\Omega b^{-d}$, and continuous $V(x,u)$, the minimum
energy configuration $u(x;v)$ is unique and smoothly varying with
$v$, except for a discrete set of shock positions where $u(x;v)$
jumps between degenerate minima.

For numerics, interfaces $u(x)$ are described by a set $I$ of
edge-sharing plaquettes $p$. Plaquettes are dual to the edges in a
regular lattice composed of $H$ layers. Each layer has $L^d$ points,
unit cell volume $\Omega_0$, and periodic BCs.  Each point is
connected to points in the layer above by $\kappa$ bonds, so that an
interface $I$ has $\kappa L^d$ plaquettes \cite{us_future}. The
energy ${\cal H}_{\rm latt}$ of $I$, confined by a well of strength
$M$ centered at $v$, is
\begin{eqnarray}
{\cal H}_{\rm latt}(v)&=&\sum_{p\in I}\left\{\frac{M^2}{2} [u(p) -
v]^2 + U(p) \right\}\,,\label{eq:H_latt}
\end{eqnarray}
where $u(p)$ is the layer index for plaquette $p$ and $U(p)$ is the
disorder potential. Long-wavelength elasticity arises from
combinatorial effects \cite{Middleton1995}. For RB disorder, $U(p)$ is
a Gaussian variable $h(p)$ with zero mean and variance $\sigma^2_0 =
1$, while for RF disorder $U(p)$ is the sum of $h(p)$ along a path of
edges connecting $p$ to the bottom layer.  RP disorder with period $P$
is constructed by stacking $H/P$ identical RB samples of thickness
$P$. Given $U(p)$, $v$, and $M$, the ground state $I_{\rm gs}$ is
found using a program that accommodates all lattices,
dimensionalities, and disorder types. The new version of the core
max-flow algorithm \cite{Middleton1995} in our code has been directly
tested against standard libraries \cite{Goldberg} and earlier energy
minimization calculations
\cite{Middleton1995,books,NohRieger2001}. The height $H$ is always
large enough that the finite size effects are controlled only by $L$
and $M$.  Lattice discreteness is evident at high values of $M$, so we
choose $M \stackrel{<}{~} 0.2$. Continuum and discrete models are then
related by equating energies ${\cal H}_{\rm latt}$ and ${\cal H}$,
displacements $u(p)$ and $u(x)$, interface areas $\Omega_0 \kappa b^d
L^d = \Omega$, well strengths $m^2=M^2(\Omega_0 b^d)^{-1}$, and
disorder strengths $\sigma=\frac{\kappa
\sigma_{0}}{2\Omega_{0}b^d}$. The effective elastic constant $K$ was
also measured \cite{us_future}.

We computed the discrete force-force correlation \cite{pld}:
\begin{eqnarray}
\Delta_{\mathrm{latt}}(v) = M^4 (\kappa L^d)
\overline{[v'+v-u_0(v'+v)][v'-u_0(v')]}, \label{deltad}
\end{eqnarray}
where the mean position $u_0(v)=(\kappa L^d)^{-1} \sum_{p\in I_{\rm
gs}} u(p)$. The averaging (overline) is for $N>10^4$ samples with
$0\le v'<P$ for RP disorder; RF and RP samples are self-averaging
over $v'$ (we slide $v'$ over more than $10^5$
times the interface width while computing minima in a window of
thickness $H\approx 20$ centered at $v'$).
The plots we present have $1\sigma$ error bars computed using direct
resampling of the data and are thus expected to overlap the large
sample number limit with a probability of $68\%$.
To check our procedure, we confirm that
$\int_0^\infty du\,\Delta_{\rm latt}(u)$ is consistent within errors
with the value $\sigma_0$ for RF disorder and with the value $0$ for
RB and RP disorders.

The FRG predicts that, for large volumes $\Omega/b^d$, the rescaled
correlator $\tilde \Delta (z)$, defined by
$\Delta(u)=m^{\epsilon-4\zeta} \tilde \Delta(u m^{\zeta})$,
converges as $m \to 0$ to the FRG fixed-point function $\tilde
\Delta^*(z)$, which depends only on $d$ and disorder.  Using
Eq.~(\ref{deltad}), convergence of
$M^{4\zeta-\epsilon}\Delta_{\rm latt}(zM^{-\zeta})$ was
evident for $L>16$ ($L > 8$ in $d=3$) and $M < 0.2$, choosing
\cite{Middleton1995} $\zeta=2/3,0.44,0.22$ for $d=1,2,3$ RB disorder
and $\zeta= (4-d) /3$ for RF disorder.  The interface widths grow
slowly ($\zeta=0$) for RP disorder.  As the FP functions still
contain an amplitude and a scale, we introduce the normalized
function $Y(z)$,
\begin{eqnarray}
\Delta(u)=\Delta(0) Y(u/\xi),  \label{split}
\end{eqnarray}
so that $Y(0)\equiv 1$ and with scale $\xi$ chosen according to
disorder type: for the RP model, $\xi=P$, for RF disorder, $\xi$ is
set so that $\int_0^\infty dz\,Y(z)=1$, and for RB disorder,
$\int_0^\infty dz\,Y^2(z)=1$. This function is predicted to be
universal with a dependence in $d$, $Y(z;d)$, that can be computed
to second order in $\epsilon:=4-d$
\cite{ChauveLeDoussalWiese2000a,LeDoussalWieseChauve2004}
\begin{eqnarray}
Y(z;d) = Y_1(z) + \epsilon Y_2(z) + O(\epsilon^2), \label{expansion}
\end{eqnarray}
with $Y_1(z)$ the 1-loop estimate \cite{DSFisher1986,BalentsFisher1993,
GiamarchiLeDoussal1995}. Computation of $K$ is required to
fix universal information
not retained in $Y(z)$, e.g., the amplitude $\Delta(0)$ for RF
disorder \cite{us_future}.

We plot illustrative examples of $Y(z)$ in Fig.\ \ref{figure1}.
In all cases, an apparently linear cusp is found for
$Y(z)$ (with finite intercepts for fits to $Y'(z)$).
The
normalized functions are remarkably close to 1-loop predictions,
with no adjustable parameter.  We now turn to a systematic analysis of
these functions, their deviation from 1-loop results, and related
data.

We start with RF disorder. The FRG predictions for the functions
$Y_1(z)$ and $Y_2(z)$ in (\ref{expansion}) are obtained from
linearizing the $O(\epsilon^2)$ relation
\cite{LeDoussalWieseChauve2004} $z = \frac{ \sqrt{Y - 1 - \ln Y -
\frac{\epsilon}{3} F(y)} }{\int_0^1 dy \sqrt{y - 1 - \ln y -
\frac{\epsilon}{3} F(y)} }$, where $F(y) = 2 y - 1 + \frac{y \ln
y}{1-y} - \frac{1}{2} \ln y + {\rm Li}_2(1-y)$. Plots of the
differences $Y(z)-Y_{1}(z)$ between the numerical result and the
1-loop prediction \cite{DSFisher1986}, for several sizes and masses in
$D=2+1$ and $D=3+1$, are shown in Fig.\ \ref{RFRBresiduals}.
There are small, but
statistically significant, systematic deviations from $Y_1(z)$.
The sign of the expected
corrections linear in $\epsilon$, $Y_2(z)$, agrees with numerics.
This function changes sign at $z_c=1.668 \ldots$, near the observed
location.
The magnitude of $\epsilon Y_2(z)$, setting $\epsilon=1$,
is nearly consistent with numerics for all $d$.
We include $0+1$ numerical results (compatible with
Refs.~\cite{pld,olaf}) for comparison.
Points for $D=2+1,3+1$ are both close to $D=0+1$ results.
Our computed slopes at the origin,
$-Y'(0)=0.815(7)$ (3+1) and $-Y'(0)=0.811(6)$ (2+1), are to be
compared with the FRG value $0.7753\ldots + (0.0328\ldots) \epsilon$
and the $d=0$ \cite{pld} value $0.8109\ldots$. The near equality of
the $d=0$ curve and $Y_2(z)$ appears to be a coincidence. Although
more work is needed to resolve the differences (e.g., $d=0$ from
$d=2,3$) the trend of the FRG results is encouraging.

For RB disorder, $R(u)$ is expected to decay (so $\Delta(u)$ has a
zero). Fixing $\xi$ as stated sets a non-universal scale.  The
differences $Y(z)-Y_1(z)$ are plotted in Fig.\ \ref{RFRBresiduals}: we
again find small but significant deviations from the 1-loop
prediction, with at most a weak dependence on $d$ (within error
bars). The $O(\epsilon^2)$ expansion in this case is found from series
and numerical solutions \cite{LeDoussalWieseChauve2004}. The resulting
$Y_2(z)$ again agrees well in sign and shape with the data, with a
magnitude given by $\epsilon\approx 1$. We have constructed 2-loop
interpolations which agree with the data in all $d$
\cite{us_future}. The situation resembles that for RF disorder, even
though deviations have the opposite sign.

Results for the function $Y(z)$ for RP disorder are shown in
Fig.\ \ref{figureRP} for $d=3$; similar results hold for $d=2$. The
1- and 2-loop FRGs predict
\cite{GiamarchiLeDoussal1995,LeDoussalWieseChauve2004} a parabolic
form, $\Delta(u)= \Delta(0) (1 - 6 u(1-u))$, as do the $d=0$
and the large-$d$ cases
(with a single shock as $m\rightarrow 0$ \cite{BalentsBouchaudMezard1996}
and many small independent shocks
per period \cite{exactRG}, respectively). Counting of
derivatives in the FRG equation has also indicated that the
parabolic form holds for any finite $d$. The parabolic form is consistent
with our results as $m\rightarrow 0$.

\begin{figure}[t]
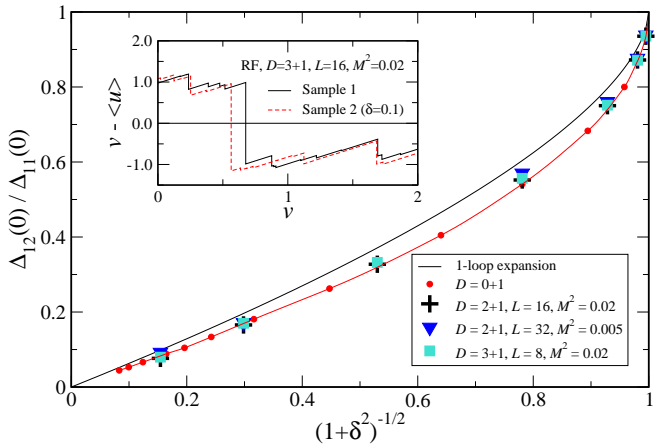
\setlength{\unitlength}{1.4mm}
\fboxsep0mm \mbox{\fig{\figsize}{chaos_inset}} \caption{A plot
of the normalized cross-correlator
$\frac{\Delta_{12}(0)}{\Delta_{11}(0)}$, computed for RF disorder in
$d=2,3$, showing the sensitivity to disorder of magnitude $\delta$,
compared to the 1-loop prediction and to numerical $D=0+1$
computations (error bars not shown; $1\sigma$ errors are about
$1/2$ of symbol size).  Inset: the renormalized pinning forces $v-u_0(v)$ for
a sample (solid line) and a $\delta=0.1$ perturbed sample (dashed
line) in a typical sample; cross-correlations of such data give the
main plot.} \label{figurechaos}
\end{figure}

The use of a harmonic well allows one to define and study the shocks in
the force landscape. As $v$ increases, sections of the manifold have
degenerate minima at positions $v_s$ and the polarization jumps
forward by $\int d^dx\,[u(x;v_s^+)-u(x;v_s^-)]$.  These are shocks in
a functional (scalar for $d=0$) decaying Burgers equation \cite{pld},
with the renormalized force $v-u(v)$ corresponding to velocity and
$m^{-1}$ corresponding to time in Burgers turbulence. Examples of
these discontinuities in the renormalized force are shown in the inset
of Fig.\ \ref{figurechaos}. We have seen shocks merge as $m$ decreases
\cite{us_future}. The moments of the renormalized force are
$S_n(v-v')=\overline{(v-v'-u(v)-u(v'))^n}$. A linear cusp in $S_2$ is
confirmed by our study of $\Delta(v)$. A prediction of the FRG in
$d>0$ \cite{pld} is that $S_3(v-v')\sim (v-v')$ at small $v-v'$, in
accord with exact results for $d=0$. Linearity of all $S_n$, $n \geq
2$ is a hallmark of intermittency in $d=0$ Burgers turbulence. Our
data show linearity of $S_3$ (Kolmogorov scaling) and $S_4$ in $v-v'$
for cases studied.  This indicates that shocks do not
cluster beyond simple statistical fluctuations.

When the pinning potential is perturbed, correlations between the
original and perturbed samples remain for RF disorder and are
described by a new chaos FRG fixed point \cite{chaospld}. We test
this prediction using related disorders $U_1(p)$ and
$U_2(p)=[U_1(p)+\delta\cdot W(p)]/\sqrt{1+\delta^2}$, where the
perturbation $W(p)$ is a mean-zero univariate Gaussian and $\delta$ is the
perturbation strength. We measured the cross correlator
$\Delta_{12}(v-v') = \kappa L^d M^4 \overline{(v- u_{0,1}(v))
(v'- u_{0,2}(v'))}$. We check the sum rule $\int_0^\infty
du\,\Delta_{12}(u)=\sigma/\sqrt{1+\delta^2}$ and normalize via
$Y_s(z)=\Delta_{12}(\xi z)/\Delta_{11}(0)$. We find
(Fig.~\ref{figure1}) that $Y_s(z)$ is rounded, as predicted
\cite{chaospld}. The computed $Y_s(0)$ is near the 1-loop
prediction (see Fig.\ \ref{figurechaos}).

Our numerical calculations confirm the main features of the FRG
approach to the glassy system of pinned interfaces, especially the
shape of $R''(u)$ and its linear cusp, for a variety of disorders and
dimensionalities.  FRG computations to 2-loop order significantly
improve upon the 1-loop results.  The functional shocks found are
consistent with expectations; their statistics merit further study.

The authors thank the KITP and the Aspen Center for Physics for their
hospitality. We acknowledge support from NSF Grants 0109164 and
0219292 (AAM) and ANR program 05-BLAN-0099-01 (PLD and KW).


\begin{thebibliography}{100}

\bibitem{HuseHenleyFisher}
D.A. Huse, C. L. Henley and D.S. Fisher, Phys.~Rev.~Lett.~{\bf 55}, 2924 (1985).

\bibitem{Johansson2000}
K. Johansson, Commun. Math. Phys. {\bf 209}, 437 (2000).

\bibitem{MP}
M. Mezard and G. Parisi, J. Phys. I {\bf 1}, 809 (1991).

\bibitem{Dominicis}
C. de Dominicis, {\em et al}, in A.P. Young, editor, {\em Spin glasses and
Random Fields}, World Scientific, Singapore, 1997.

\bibitem{NattermannBookYoung}
T.~Nattermann, and C. Belanger, {\em Ibid}.

\bibitem{DSFisher1986}
D.S. Fisher, Phys. Rev. Lett. {\bf 56} 1964 (1986) and Phys. Rev. B
{\bf 31}, 7233 (1985).

\bibitem{reviews_pinning}
G.~Blatter, {\em et al}, Rev. Mod. Phys. {\bf 66}, 1125 (1994);
T.~Nattermann and S.~Scheidl, Advances in Physics {\bf 49} (2000)
607.

\bibitem{GiamarchiLeDoussal1995}
T.~Giamarchi and P.~Le Doussal, Phys. Rev. B {\bf 52}, 1242 (1995).

\bibitem{depinning}
O. Narayan and D.S. Fisher, Phys. Rev. B {\bf 48}, 7030 (1993).
Nattermann, {\em et al}, J. Phys. II (France) {\bf 2}, 1483 (1992).

\bibitem{creep}
P.~Chauve, T.~Giamarchi, P.~Le Doussal, Phys. Rev. B {\bf 62}, 6241 (2000).

\bibitem{quantum}
L. Balents, Europhys. Lett. {\bf 24}, 489 (1993).

\bibitem{chaospld}
P. Le Doussal, cond-mat/0505679.

\bibitem{FRGfermions}
C.J. Halboth and W. Metzner, Phys. Rev. B {\bf 61}, 7364 (2000).

\bibitem{BalentsFisher1993}
L.~Balents and D.S. Fisher, Phys. Rev. B {\bf 48}, 5949 (1993).

\bibitem{BalentsBouchaudMezard1996}
L. Balents and J.P. Bouchaud and M. M\'ezard,
J. Phys. I (France) {\bf 6}, 1007 (1996).

\bibitem{Middleton1995}
A.A. Middleton, Phys. Rev. E {\bf 52}, R3337 (1995); D.~McNamara,
A.A. Middleton  and C. Zeng, Phys. Rev. B {\bf 60}, 10062 (1999).

\bibitem{books} see e.g.
M.J. Alava, P.M. Duxbury, C. Moukarzel and H. Rieger in {\it Phase
transitions and critical phenomena}, Ed. C. Domb and J. L. Lebowitz,
Academic Press, San Diego, {\bf 18}, 141 (2001).

\bibitem{NohRieger2001}
J.D. Noh and H.~Rieger, Phys. Rev. Lett. {\bf 87}, 176102 (2001).

\bibitem{ChauveLeDoussalWiese2000a}
P.~Chauve, P.~Le Doussal  and K.~Wiese, Phys. Rev. Lett. {\bf 86},
1785 (2001).

\bibitem{exactRG}
L. Balents and P. Le Doussal, Annals of Physics, {\bf 315}, 213
(2005); P.~Chauve and P.~Le Doussal, Phys. Rev. E {\bf 64}, 051102
(2001); S. Scheidl, Y. Dincer cond-mat/0006048.

\bibitem{LeDoussalWieseChauve2004}
P.~Le Doussal, K.J. Wiese and P.~Chauve, Phys.Rev. E {\bf 69},
026112 (2004); P. Le Doussal, K. J. Wiese, Phys. Rev. E {\bf 72},
035101(R) (2005),
Phys. Rev. Lett. {\bf 89} 125702 (2002).

\bibitem{LeDoussalWieseChauve2002}
P. Le~Doussal, K.J. Wiese and P. Chauve, Phys. Rev. B {\bf
66}, 174201 (2002).

\bibitem{pld}
P. Le Doussal, cond-mat/0605490 and in preparation.

\bibitem{Binder}
K. Binder, Z. Physik B {\bf 43}, 119 (1982); E. Brezin and J.
Zinn-Justin, J. Nucl. Phys. B {\bf 257}, 867 (1985); T.W.
Burkhardt, B. Derrida, Phys. Rev. B {\bf 32}, 7273 (1985).

\bibitem{us_future}
P. Le Doussal, A. Middleton, K. J. Wiese, to be published, and
a longer work, in preparation.

\bibitem{Goldberg}
A. V. Goldberg, Optimization Library, available at
http://www.avglab.com/andrew/.

\bibitem{olaf}
O.~Duemmer, to be published.


\end{thebibliography}
\end{document}